# Structure, dynamics and light localization in self-induced plasma photonic lattices

Rotem Kupfer, Boris D. Barmashenko, and Ilana Bar[1]

*Department of Physics, Ben-Gurion University of the Negev, Beer-Sheva 84105, Israel*

Detailed two dimensional particle-in-cell (PIC) simulations and numerical calculations of electron density profiles, based on a simplified model, were performed to show for the first time that underdense plasma, induced by two pairs of counterpropagating femtosecond-laser pulses in a gas, could be manipulated by ponderomotive-optical lattices to form periodic structures. These novel phenomena facilitate the localization and enhancement of the generating laser pulse intensities by the self-induced plasma photonic lattices (PPLs) and exhibit unique spatiotemporal dynamics. Variation of the initial plasma density profile and the configuration of the interacting pulses enabled control over the attainable PPL structures. It is predicted that by using a non-uniform initial plasma density, light emission in a preferred direction could be obtained.

PACS numbers: 42.65.Sf, 52.38.-r, 52.65.Rr

---

[1] Email address: ibar@bgu.ac.il

# I. INTRODUCTION

Localization and confinement of light into wavelength scale cavities is of great interest in many fields of physics and engineering [1] [2]. Indeed, methods using different optical structures, like metallic mirrors, photonic band-gap materials [3], highly disordered media [2] and periodic photonic crystals (PCs) [4] have been suggested for their ability to control light in unusual and compelling ways. For example, the PC [1] is a dielectric patterned material, containing air holes, with a spatially varying but three-dimensional (3D), periodic structure. Taking advantage of the design flexibility and the possibility to introduce defects in otherwise perfectly PCs enabled experimental demonstration [5] [6] [7] and computational and analytical prediction [8] of light confinement in a variety of microcavities.

In particular, a microscopic counterpart of a Fabry-Perot cavity was created by introducing a defect in an otherwise perfectly periodic PC, namely a novel optical nanostructure with a periodic refractive-index modulation [5]. In addition, a single defect, created artificially inside a two-dimensional (2D) photonic bandgap structure, enabled trapping of photons propagating through a linear waveguide, which were then emitted to free space [6]. Furthermore, a silicon-based 2D PC slab, allowed fabrication of a high quality factor and small modal volume nanocavity that was used for light confinement [7].

Nevertheless, these approaches require specific devices, or prefabricated solid PCs, which are not always easy to produce and suffer from the drawback that the incident pulse intensities should be limited by the damage threshold of the involved materials. Therefore, finding variant structures that provide new ways for confining light would be beneficial. Especially, it would be interesting to test whether a new kind of self-generated plasma structures, which are practically not limited by the exciting pulse intensities, could be leveraged for this purpose.

It is well known that when an intense femtosecond (fs) laser pulse is focused in a gas, it ionizes the molecules and atoms in its path, leading to plasma channel formation [9]. The ionization occurs at the leading edge of the pulse, so that essentially the pulse propagates inside its self-generated channel [10]. By studying the nonlinear interaction of a single pulse with the plasma channel [11] [12], as well as the intersection of two incident pulses [13] [14] it was revealed that the ponderomotive force (a nonlinear drift force that pushes the charged particles toward weaker electric amplitude regions [15]) plays a crucial role in these interactions [16]. The ponderomotive force is given by:

$$\boldsymbol{F} = -q^2/(4m\gamma\omega^2)\nabla \mathrm{E}^2 \quad , \quad (1)$$

being proportional to the time averaged intensity gradient, where E and $\omega$ are the time independent amplitude and angular frequency of the driving electric field, respectively, $q$ and $m$ are the charge and mass of the particles, respectively, and $\gamma$ is the Lorentz factor. It has been recently shown that this force allows plasma density manipulations, while utilizing the interference pattern of two [17] [18] [19], or three [20] laser pulses. Specifically, we have shown that by exploiting the Moiré patterns of the interfering electric fields, plasma gratings can be produced [19], which in fact might act similarly to photonic waveguide arrays.

Hence, we suggest taking this approach one step further and finding out whether it would be possible to use a ponderomotive-optical lattice, generated by two pairs of counterpropagating and intersecting fs laser pulses, to form self-generated plasma photonic lattices (PPLs). The external laser beams, ionize the gas to generate the PPLs, thus producing refractive indices with lattice structures that last through the duration of the creating/maintaining laser pulses. In particular, we are interested in finding under which conditions these PPLs, could be generated and to obtain a mechanistic insight regarding their structures, dynamics and their possibilities of light confinement and manipulations.

## II. METHODS

Briefly, the simulations were performed using our finite-difference time-domain particle-in-cell (FDTD-PIC) code, which has been used to explain previous experimental results and to predict others [19]. The kinetic PIC method is a well-known technique [21] [22] [23] for modeling laser-plasma interactions as well as space plasmas. For the numerical computation, a discrete grid of electromagnetic fields and a finite number of macro-particles were assumed. Each particle represented a macroscopic number of real electrons with proper mass and charge that could move continuously inside the electromagnetic mesh. The ions were assumed to be immobile on the time scale of interest, as justified by the results of several executions with mobile ions. Because of its simplicity, PIC inherently includes all nonlinearities and can be considered as a first principle method [24] [25]. The only limitations were computational rather than physical, i.e., the model was restricted to 2D transverse electromagnetic waves in the z direction

(TEz, requiring the polarization of incident waves to be in-plane) and to a finite number of macro particles (~12 x $10^6$).

For numerically simulating the laser plasma interaction, the FDTD-PIC algorithm worked according to the numerical scheme described in Ref. [19]. First, the particles and fields were initialized and afterwards, at each time step, the Maxwell's curl equations were solved, using Yee's algorithm [26] at each grid point. Then, the electromagnetic fields were interpolated to the exact position of each particle and the corresponding Lorentz force was calculated. Each particle was pushed using a second order accurate scheme [27] and the current density caused by this motion was calculated using a charge conserving method [28] and injected back into the curl equations. According to this scheme, Maxwell's time dependent equations, the equations of motion with Lorentz force and the expression for the local current density form a self-consistent system. While the algorithm was executed serially, the calculations of each step were performed using parallel computer unified device architecture (CUDA) on a graphic processing unit (GPU), requiring only modest computational resources. The simulation area was surrounded by a perfectly matched layer [29] to avoid reflection from the edges. The pulses were injected at the grid edge using the total-field/scattering-field (TF-SF) scheme [23] for which pulse wavelength, intensity, duration and waist could be controlled. In addition, the focusing conditions for the pulses and their incident angles could be adjusted by introducing proper spatially dependent phases at the edge. At each time step, the local field at each point in the simulation area, as well as the speed and location of each particle could be examined. The spectrum of the electromagnetic field at any point on the grid was calculated by recording the instantaneous values of the local electric field, followed by Fourier transform (FT).

Two pairs of counterpropagating incident transform limited Gaussians laser pulses, with central wavelengths of 1 μm and 50 fs duration (peak intensities of 2.5 x $10^{16}$ W/cm$^2$ [30] and waist of 4 μm), propagating in underdense air plasma, were simulated. Each counterpropagating pulse pair had similar polarization, while the horizontal and vertical propagating pulses had perpendicular polarization with respect to each other. The electric field mesh size, $\Delta x$, was 20 x 20 nm with 12 particles per cell and a 22 attoseconds step size, $\Delta t$, stricter than the Courant condition for a 2D mesh [23], i.e., $\Delta t = \Delta x/3c$, where $c$ is the light velocity. It was found that under these conditions, the code was numerically stable for extremely long executions (> 3 x $10^6$ iterations, much longer than in the simulations reported here). Further refinement of the grid or increase in the number of particles per cell showed no change in the calculation results.

## III. RESULTS AND DISCUSSION

### A. Generation, structure and dynamics of plasma photonic lattices

A simulation snapshot of the electric field intensity (red solid surface) and plasma density (blue meshed surface), during the interaction of the pulse peak intensities ($t = 0$ fs) on the same spatial coordinates, is shown in Fig. 1. It can be clearly seen that as a result of the intersection of the four incident pulses at the center, an interference pattern of the electric field occurred. Like in two laser pulse intersection [19], this interference pattern applied a ponderomotive force, causing condensation of the initially uniform plasma into optical lattice nodes and consequently leading to plasma peaks formation, i.e. a PPL.

The temporal evolution of this lattice in the intersection region is presented in Fig. 2, but it is best understood by the movie in the supplementary material [31]. While the electron density is initially uniform ($n_e$ = 1 x 10$^{20}$ cm$^{-3}$), as the pulse intensities increase, a plasma lattice starts to build up already at -40 fs (for a pulse width of 50 fs), implying that the intensity available at this time (~10$^{14}$ W/cm$^2$) is high enough to initiate plasma displacement. The prominence of the plasma lattice increases until the laser peak intensities interact (0 fs) and fades later. Furthermore, the plasma density peaks are formed at the nodes of the standing electric wave (the lattice period is $\lambda/2$) and around the pulse peak intensities half of them disperse, resulting in a lattice constant of $\lambda$ and plasma peak densities of 7 x 10$^{20}$ cm$^{-3}$, slightly below the critical plasma density, $n_c = m_e \omega_L^2 / 4\pi e^2$ [32], where $m_e$ and $e$ are electron mass and charge and $\omega_L$ is the laser frequency.

Essentially, the plasma lattice is composed of two sub-lattices, related to spatially narrow and broad crests, evolving in time [see Fig. 2(b)]. In fact, by keeping track of this evolution, it is seen that the neighboring plasma peaks of the two sub-lattices experience different dynamics and at the peak intensity of the interaction, Fig. 2(c), the lattice of broad crests already disappeared. This behavior could be attributed to the relative angle between the vectors of the electric field and of the ponderomotive force, presented in Fig. 3, caused by their different spatial periodicity. While the stable sub-lattice (narrow crests) is characterized by the action of parallel or anti-parallel electric and ponderomotive vectors, the unstable one (broad crests) exhibits an angle between them. Hence, the total force applies a torque on the plasma peaks, rotating them by 90° in each optical lattice cycle [31]. In the next cycle the electric field vector alters its direction, imposing an opposite torque. For the polarizations used in our simulation, the vectors of the electric, $F_e$, and ponderomotive, $F_p$, forces, acting on the electrons, scale as:

$$F_e \sim -\hat{y}cos(k_x x) - \hat{x}cos(k_y y) \quad ,(2)$$

$$F_p \sim \hat{y}sin(2k_x x) + \hat{x}sin(2k_y y) ,(3)$$

respectively, where $k_x$ and $k_y$ are the wavevectors. In principle, it is reasonable to assume that by choosing different polarizations, the lattice peak dynamics could be altered due to different ponderomotive and electric force morphology (although we were not able to simulate different polarizations, due to the 2D computational restrictions).

We note that for the case of an optical lattice (standing wave with time dependent envelope), higher orders of the ponderomotive force [33] result in higher frequency periodic corrections, which could explain the dispersion of the unstable sub-lattice once the higher derivatives of the envelope switch sign. As can be seen from Fig. 2 and from the movie in the supplementary material [30], the lattice-modulated plasma lasts only for a very short time, depending on the duration of the creating/maintaining laser pulses. Nevertheless, by simulating external pulses of other shapes, i.e., flat-top envelopes, we have found that the PPL could be maintained for much longer durations for intensities of the order of $\sim 10^{14}$ W/cm². In this case, the dynamics of the two sub-lattices disappeared and the plasma remained in a stable $\lambda$ period lattice for over a few tens of ps.

Some physical insight regarding the intriguing phenomena observed here can be alternatively derived from the following simplified model. Specifically, balancing of the plasma pressure gradient force, $F_{pg} = -T_e \nabla n$, and the ponderomotive force, Eq. (1) [34] [35] leads to:

$$n_e^{-1} \nabla n = -\frac{e^2 \nabla E^2}{m_e \omega_L^2 T_e}, \quad (4)$$

where $T_e$ is the temperature (in eV, assumed to be constant) and $\omega_L$ is the laser frequency. The electric field amplitude (including the temporal envelope, while neglecting the transverse beam profile) is:

$$E(x,y,t)^2 = E_0^2 \times exp\left(-\frac{2t^2}{\tau_p^2}\right) \times [cos^2(kx) + cos^2(ky)], \quad (5)$$

where $2\tau_p$ is the pulse duration and $k$ is the laser wavevector. Therefore, the periodic spatial plasma profile is obtained by integration to give the electron density:

$$n_e(x,y,t) = n_{e,0}(t) exp\left(-\frac{e^2 E(x,y,t)^2}{m_e \omega_L^2 T_e}\right), \quad (6)$$

where the prefactor $n_{e,0}$ should be determined by conservation of the number of electrons:

$$n_{e,0}(t) = n_0 V \times \left(\iiint exp\left(-\frac{e^2 E(x,y,t)^2}{m_e \omega_L^2 T_e}\right) dV\right)^{-1}, \quad (7)$$

where $n_0V$ is the initial electron number in the volume. The electron density profile calculated numerically, using Eq. (6), for a pulse wavelength of 1 μm at peak intensity of 2 x 10$^{16}$ W/cm$^2$, initial plasma density of 1 x 10$^{20}$ cm$^{-3}$ and electron temperature of 800 keV is presented in Fig. 4. It can be clearly seen that this model predicts a PPL that qualitatively resembles the one obtained by the PIC simulation, Fig. 2(c), for similar conditions. Nevertheless, since this model does not consider the dynamics and evolution of the two different sub-lattices, which were obtained by the PIC simulation, it actually does not account for the disappearance of the unstable one.

**B. Light localization in optically induced plasma photonic lattice**

Examination of the electric field intensity of the four pulses in the intersection region showed that using higher initial plasma densities (higher air pressure) allowed field localization at the intersection point [Fig. 5]. In particular, for a pulse intensity of 2.5 x 10$^{16}$ W/cm$^2$ and a plasma density of ~2 x 10$^{19}$ cm$^{-3}$ (in atmospheric air [12] [30]) the peak intensity in the interference pattern was the sum of the four pulse intensities (10$^{17}$ W/cm$^2$) [Fig. 5(a)]. However, at similar laser pulse intensities, but initial plasma density of 1 x 10$^{20}$ cm$^{-3}$, the light became confined to the interference pattern center, leading to a 70 % enhancement, on account of the intensity at larger distances from the center. This behavior is even more pronounced for the higher plasma density of 9 x 10$^{20}$ cm$^{-3}$ [Fig. 5 (b), just below $n_c$] that led to intensity enhancement of 350 %. Furthermore, the horizontal cut of the spatial intensity distributions for pulses propagating in dense air 9 x 10$^{20}$ cm$^{-3}$ (red solid line) and in vacuum (blue circle line), Fig. 6, shows the intensity accumulation at the center of the PPL, on the expense of the intensity in the outer region.

This light localization can be attributed to confinement of the incident laser pulse intensities by refractive index variations in the PPL, similar to those occurring in PCs. The time dependent local refractive index of the plasma is related to $n_e(x,y,t)$ and to $n_c$ by [36]:

$$n(x,y,t) = \sqrt{1 - n_e(x,y,t)/n_c} \quad . (8)$$

In spite of the unique temporal dynamics the plasma peaks still remain well located in the optical lattice nodes, leading to the PPL formation. Therefore, the four pulses essentially act as pump pulses for the stable electromagnetic mode at the self-induced PPL center. It is worth emphasizing that no new frequencies appeared in the FT spectrum of the field, implying that the light localization occurred due to a linear process, rather than nonlinear ones.

In fact for higher initial densities, the plasma peaks become higher than $n_c$, leading to enhanced scattering and light confinement. Since the maximal value of the initial plasma density that allows localization is bound by $n_c$ and depends on the incident wavelength, it is found that the optimal enhancements occur at initial densities, slightly below $n_c$. Particularly, optimal intensity enhancements were obtained for laser pulse wavelengths of 1 μm and 800 nm at plasma densities of 9 x $10^{20}$ and 1.6 x $10^{21}$ cm$^{-3}$, respectively. The temporal behavior of the electric field intensity for the former is presented in a movie attached in the supplementary material [37]. As can be seen, the localization is caused by incident pulse self-action and pumping through the PPL, contrary to that occurring in a predetermined PC with perpendicular or near field pumping.

It is suggested that a qualitative explanation for this localization can be derived from a simplified model, considering the presence of four effective distributed Bragg reflectors (DBRs) [38] for the specific configuration encountered here, i.e., four mirrors around an optical cavity. Hence, when a laser pulse aims to exit the PPL structure, a fraction of it, propagating between the plasma peaks, is unaffected, while the other portion experiences the existence of the effective DBRs. The reflection coefficient for a single DBR is roughly approximated as [39]:

$$R(t) = [\frac{(\max{(n(t))})^{2N} - (\min{(n(t))})^{2N}}{(\max{(n(t))})^{2N} + (\min{(n(t))})^{2N}}]^2, \quad (9)$$

where $N$ is the ratio between the lattice length and the period and $n(t)$ is the time dependent refractive index as calculated from Eqs. (6) and (8). The temporal evolution of the reflection coefficient of a DBR for several values of initial densities is presented in Fig. 7. It can be clearly seen that the reflection coefficients for all initial plasma densities evolve like the temporal envelope of the pulses, resulting in larger light confinement at the peak intensities and thus agreeing well with the observation that the utmost light localization in the PIC simulations was obtained for the highest plasma density, see Fig. 5(b). However, since the PPL structure is tailored for specific wavelength and phase of the creating pulses, this idealized approximation cannot lead to quantitive agreement with the PIC results.

## C. Manipulation of the light localization, emission and plasma photonic lattice structure

Up to now, it was shown that the stable mode position is in the center of the PPL, Fig. 5, where the four pulses intersect. We conjectured that this mode can be displaced by manipulating the initial plasma density and consequently the local refractive index [40] [41]. For example this kind of displacement could be obtained by altering the initial density profile, see Fig. 8(a). Here, a Gaussian area with peak density of $7.5 \times 10^{20}$ cm$^{-3}$ was placed on top of the uniform density of $5 \times 10^{20}$ cm$^{-3}$. This perturbation interfered with the lattice formation in the corresponding region [Fig. 8(b)], causing a shift of the mode from the simulation center toward a position, further away from the perturbation [Fig. 8(c)]. In contrast, by using a density dip, the stable mode could be attracted from its position toward the center.

In addition, by introducing a gradient in the initial plasma density (a linear ramp with a $3 \times 10^{23}$ cm$^{-4}$ slope over a 10 μm length for a $5 \times 10^{20}$ cm$^{-3}$ origin density) the symmetry of the plasma lattice was broken, leading to a yet more practical outcome. In this case the emission lost its symmetry and the pulse that propagated along the plasma density gradient experienced a different refractive index than the other three pulses. This resulted in amplification of the former pulse at the expense of the other three [Fig. 9(a)], together with an intensity enhancement of over 90 % at the center. Furthermore, Fig. 9(b) clearly shows that the amplified pulse (blue solid line) and its counter-propagating (red dashed line) experience a relative phase shift, while the perpendicular couple (black circles and green triangles) not, reflecting the variation in refractive indices in the different directions, due to the plasma density gradient. This leads us to consider again the outcome of the simplified model of the DBRs and to suggest a qualitative explanation for the directed emission. It seems that by introducing a plasma density gradient in a specific direction, the pulse across this path encounters a lower *R(t)* and hence enhanced emission.

This kind of plasma profiles could be obtained by manipulating the initial gas density, probably by the use of a sharp edge jet and colloids of liquid phase particles in the gas. A possible application of the PPL could be related to ion lasing [42]. It is supposed that by combining a stationary (on our timescale) high density ionic lasing medium, with a cavity caused by the self-induced PPL, a new kind of plasma based laser could be realized, while exploiting higher plasma densities and pumping intensities.

In addition to the plasma density manipulations, we pretended that control over the plasma lattice structure could be achieved. To test this option, calculations based on the simplified model of Sect. IIIA (Eq. 6) for the low intensity regime (~$10^{14}$ W/cm$^2$) were performed. This intensity was chosen since in this regime the dynamics is absent, and the results of the simplified model and PIC simulations agree well. The predicted PPL structures were calculated by considering the different configurations employed for obtaining the optical lattices in Refs. [43] [44] [45]. One configuration included two couples of counterpropagating pulses with similar polarizations (1 and 2) and another couple of counterpropagating pulses with perpendicular polarization (3) [Fig. 10(a)] and opposite phase:

$$E(x,y,t)^2 = E_0^2 \times exp\left(-\frac{2t^2}{\tau_p^2}\right) \times [A_1^2 cos^2(kx) + A_2^2 cos^2(ky) + A_3^2 sin^2(ky) + 2A_1 A_2 cos(kx) cos(ky)], \quad (10)$$

where $A_1$, $A_2$, and $A_3$ are the amplitudes of the electric fields of the different couples of pulses, respectively. Then, by precisely controlling the amplitude of each couple, various structures were obtained. For ratios of 1/4/2, 1/8/1 and 1/4/4 between the respective couples, structures of honeycomb [Fig. 10 (c)], chains [Fig. 10(d)] and dimmer lattice [Fig. 10(e)], respectively, were obtained. Furthermore, by using a second configuration, i.e., a triangular optical lattice [Fig. 10 (b)], a Kagome structure was obtained [Fig. 10(f)]. It should be admitted that at this stage, PIC simulation of these PPL structures and their dynamics could not be exerted because of the polarization limitation of our code.

## IV. CONCLUSIONS

In this work, we have introduced the concept of generating PPLs, using ponderomotive-optical lattices. These unique structures are characterized by interesting spatio-temporal dynamics of two sub-lattices, maintained by the electric and ponderomotive force morphology. A specific periodic plasma structure results in periodic refractive index variations that may support a stable enhanced electromagnetic mode at the intersection point, which is significantly enhanced, by a self-action effect, compared to its intensity in vacuum. Although the lifetime of the PPLs is limited by the generating pulses duration, it has been shown that it can be maintained for longer terms by low intensity, flat-top pulses.

An alternative simplified model was derived to verify the feasibility of creating PPLs, to qualitatively explain the light localization and to suggest how other structures could be obtained. In addition, it was shown that a plasma density gradient can lead to enhanced emission, out of the PPL, in a preferred direction. Although further studies are required, it is proposed that these phenomena could possibly be exploited for new types of plasma based optical devices and lasers.

We greatly acknowledge the support of the James Franck Binational German-Israeli Program in laser-matter interaction.

# Figure captions

FIG.1. (Color online) A simulation snapshot, visualizing the instantaneous electric field intensity (red solid surface) and plasma density (blue meshed surface), during the interaction of the incident lasers pulses at t = 0 fs (peak intensities of the incident pulses). The four pulses intersect at the center generating an interference pattern with the plasma trapped at the ponderomotive-optical lattice nodes.

FIG. 2. (Color online) Plasma density profile in the intersection region at different times, relative to the laser pulses onset, for a uniform initial plasma density of 1 x $10^{20}$ cm$^{-3}$. A plasma lattice with a lattice constant of $\lambda/2$ begins to form prior to the pulse peak intensities interaction, leading to maximal density crests at 0 fs. Around this time, half of the plasma peaks disperse, leaving a lattice with a $\lambda$ period. Following the interaction of the pulses the plasma gradually returns to uniform density.

FIG. 3. (Color online) Electric and ponderomotive force vectors (normalized to the same scale) at the peak of the optical-lattice period. Neighboring ponderomotive lattice nodes differ in angles between the vectors of the electric and ponderomotive forces, resulting in two distinctive plasma sub-lattices.

FIG. 4. (Color online) Electron density profile calculated numerically using eq. (6) for laser wavelength of 1 μm at the peak intensity of 2 x $10^{16}$ W/cm$^2$, initial plasma density of 1 x $10^{20}$ cm$^{-3}$ and electron temperature of 800 keV.

FIG. 5. (Color online) Instantaneous peak electric field intensities, obtained, in the intersection region of the laser pulses for a wavelength of 1 μm at an initial plasma density of (a) $2 \times 10^{19}$ cm$^{-3}$ and (b) $9 \times 10^{20}$ cm$^{-3}$.

FIG. 6. (Color online) Instantaneous spatial intensity distribution (log scale) for pulses at a wavelength of 1 μm and peak intensity of $2 \times 10^{16}$ W/cm$^2$, propagating in an initial plasma density of $9 \times 10^{20}$ cm$^{-3}$ (red solid line) and in vacuum (blue circle line), showing the intensity accumulation at the lattice center on behalf of the intensity in the outer region

FIG. 7. (Color online) Approximated reflection coefficient ($R$) as a function of time for a distributed Bragg reflector at different initial plasma densities.

FIG. 8 (Color online) The effect of a perturbation in the initial plasma density on the localization position: (a) a Gaussian density perturbation with peak density of $7.5 \times 10^{20}$ cm$^{-3}$ was placed on top of (b) a uniform plasma density of $5 \times 10^{20}$ cm$^{-3}$. The perturbation prevents the plasma lattice to be fully formed in this region and (c) results in shifting of the stable electromagnetic mode from the perturbation region.

FIG. 9. (Color online) The effect of plasma density gradient on the time integrated emission of the plasma lattice: (a) spatial intensity distribution and (b) a horizontal cut of the integrated intensity of the four pulses as they propagate from the cavity center: The amplified pulse (blue solid line, propagates along the gradient) its counter-propagating (red dashed line) and the perpendicular couple, (black circles and green triangles)

FIG. 10. (Color online) Schematic of the pulse configurations used for obtaining the plasma photonic lattices: (a) six pulses with their polarizations and (b) three pulses. Arbitrary plasma lattice structures as calculated numerically using Eq. (6) for a laser wavelength of 1 μm at peak intensity of $1 \times 10^{14}$ W/cm$^2$, initial plasma density of $1 \times 10^{19}$ cm$^{-3}$ and electron temperature of 40 keV. The structures (c) honeycomb, (d) chains and (e) dimmer lattice were obtained using the (a) configuration and the (f) Kagome lattice by the (b) configuration.

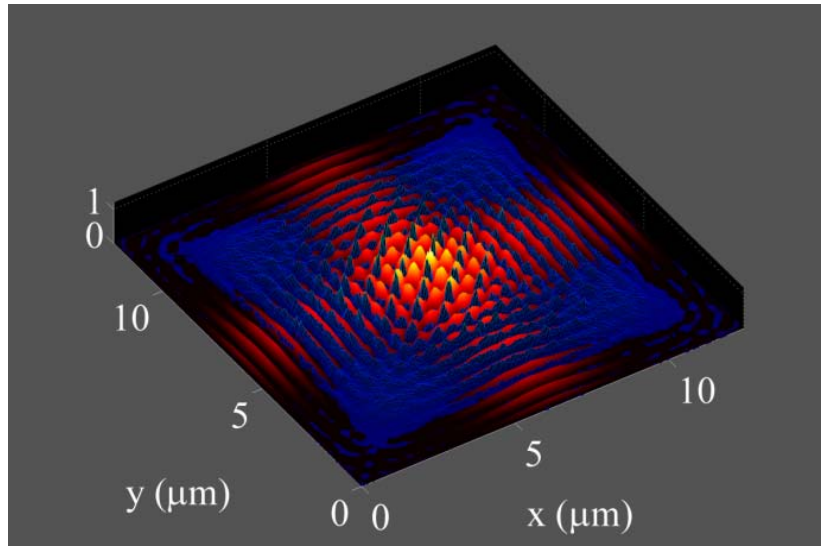

FIG.1. (Color online) A simulation snapshot, visualizing the instantaneous electric field intensity (red solid surface) and plasma density (blue meshed surface), during the interaction of the incident lasers pulses at t = 0 fs (peak intensities of the incident pulses). The four pulses intersect at the center generating an interference pattern with the plasma trapped at the ponderomotive-optical lattice nodes.

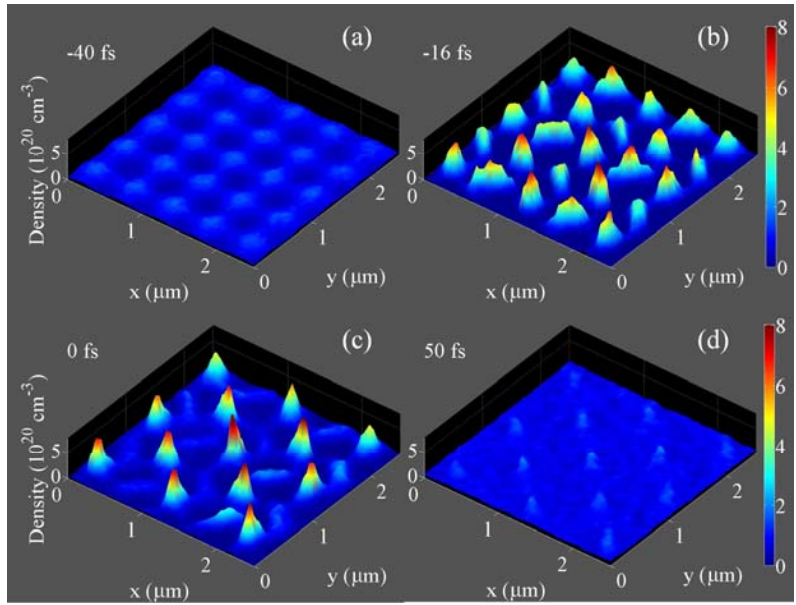

FIG. 2. (Color online) Plasma density profile in the intersection region at different times, relative to the laser pulses onset, for a uniform initial plasma density of 1 x $10^{20}$ cm$^{-3}$. A plasma lattice with a lattice constant of $\lambda/2$ begins to form prior to the pulse peak intensities interaction, leading to maximal density crests at 0 fs. Around this time, half of the plasma peaks disperse, leaving a lattice with a $\lambda$ period. Following the interaction of the pulses the plasma gradually returns to uniform density.

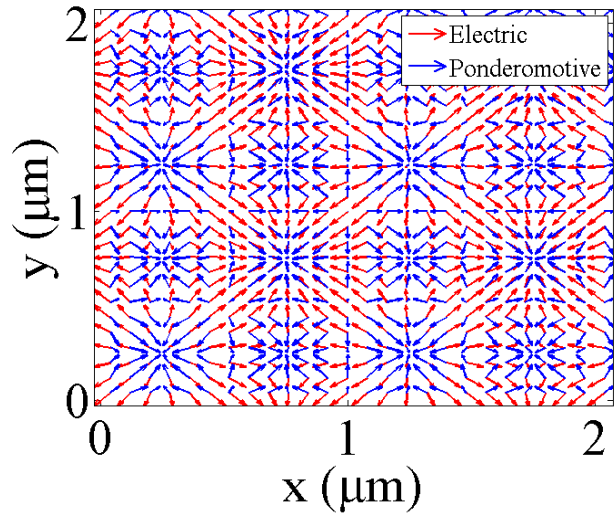

FIG. 3. (Color online) Electric and ponderomotive force vectors (normalized to the same scale) at the peak of the optical-lattice period. Neighboring ponderomotive lattice nodes differ in angles between the vectors of the electric and ponderomotive forces, resulting in two distinctive plasma sub-lattices.

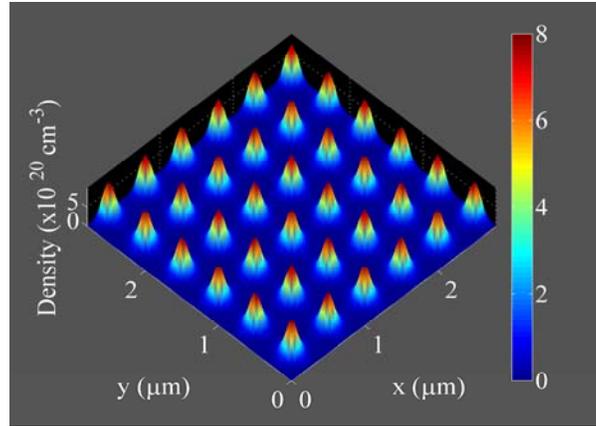
FIG. 4. (Color online) Electron density profile calculated numerically using eq. (6) for laser wavelength of 1 μm at the peak intensity of 2 x $10^{16}$ W/cm$^2$, initial plasma density of 1 x $10^{20}$ cm$^{-3}$ and electron temperature of 800 keV.

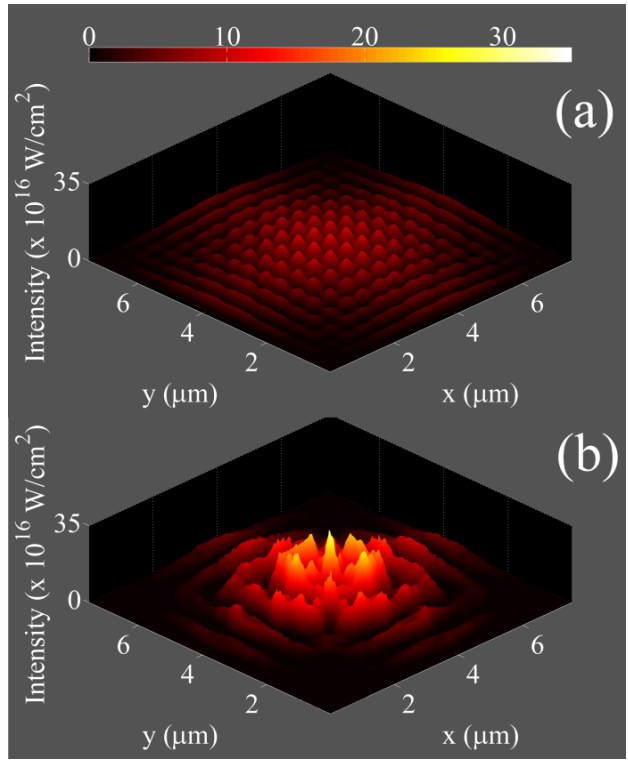

FIG. 5. (Color online) Instantaneous peak electric field intensities, obtained, in the intersection region of the laser pulses for a wavelength of 1 μm at an initial plasma density of (a) $2 \times 10^{19}$ cm$^{-3}$ and (b) $9 \times 10^{20}$ cm$^{-3}$.

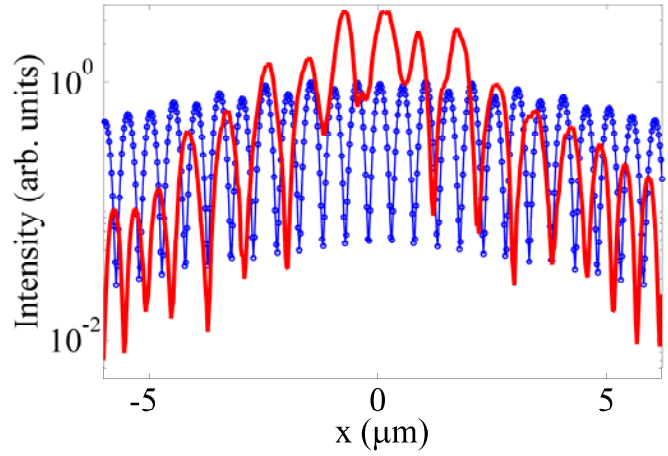

FIG. 6. (Color online) Instantaneous spatial intensity distribution (log scale) for pulses at a wavelength of 1 μm and peak intensity of 2 x $10^{16}$ W/cm$^2$, propagating in an initial plasma density of 9 x $10^{20}$ cm$^{-3}$ (red solid line) and in vacuum (blue circle line), showing the intensity accumulation at the lattice center on behalf of the intensity in the outer region

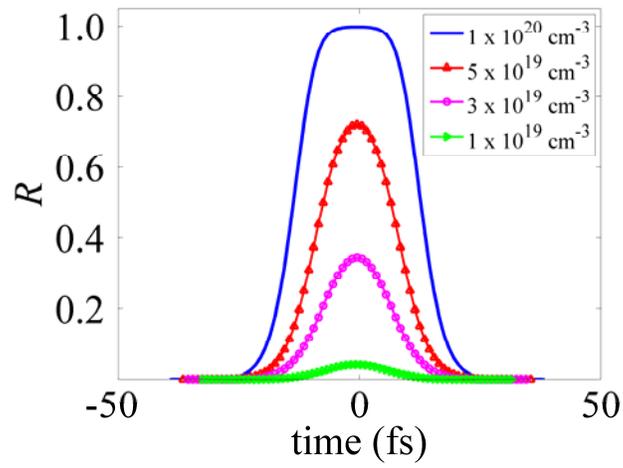

FIG. 7. (Color online) Approximated reflection coefficient ($R$) as a function of time for a distributed Bragg reflector at different initial plasma densities.

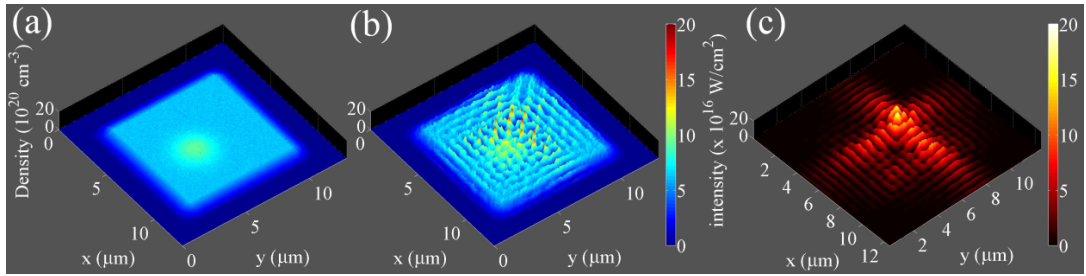

FIG. 8 (Color online) The effect of a perturbation in the initial plasma density on the localization position: (a) a Gaussian density perturbation with peak density of 7.5 x $10^{20}$ $cm^{-3}$ was placed on top of (b) a uniform plasma density of 5 x $10^{20}$ $cm^{-3}$. The perturbation prevents the plasma lattice to be fully formed in this region and (c) results in shifting of the stable electromagnetic mode from the perturbation region.

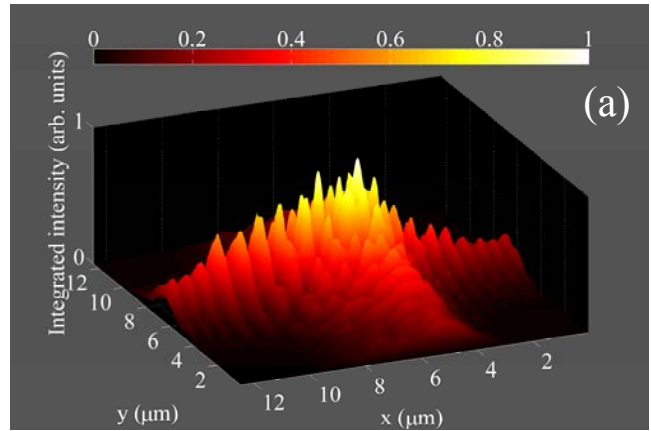

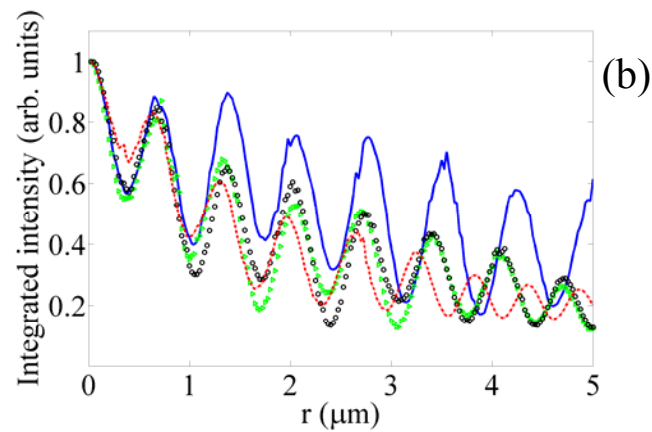

FIG. 9. (Color online) The effect of plasma density gradient on the time integrated emission of the plasma lattice: (a) spatial intensity distribution and (b) a horizontal cut of the integrated intensity of the four pulses as they propagate from the cavity center: The amplified pulse (blue solid line, propagates along the gradient) its counter-propagating (red dashed line) and the perpendicular couple, (black circles and green triangles)

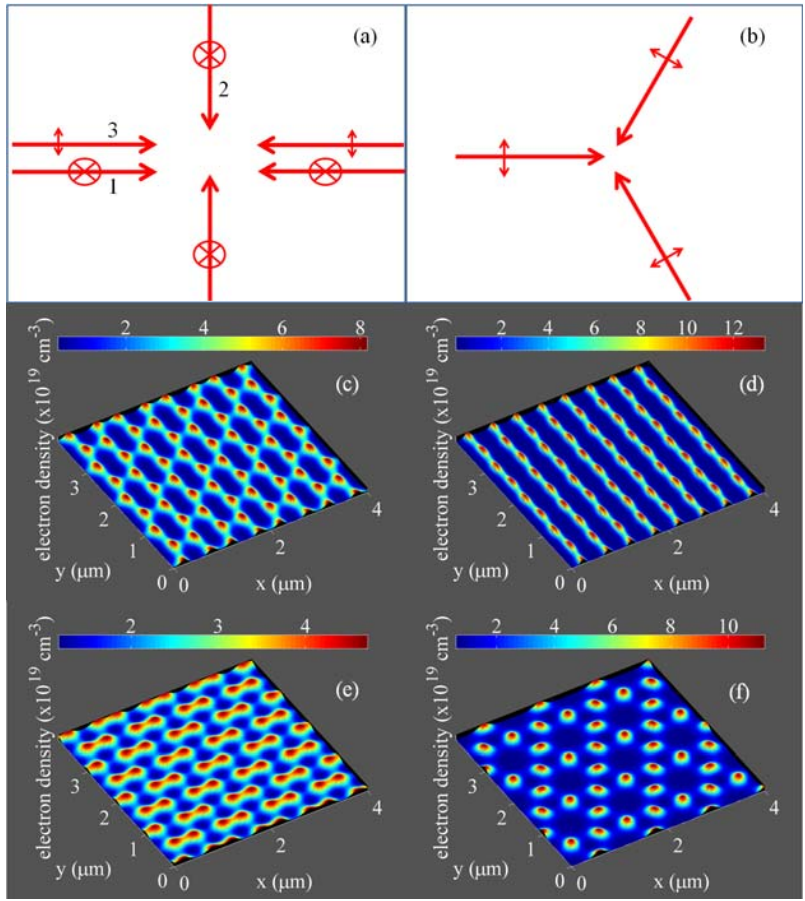

FIG. 10. (Color online) Schematic of the pulse configurations used for obtaining the plasma photonic lattices: (a) six pulses with their polarizations and (b) three pulses. Arbitrary plasma lattice structures as calculated numerically using Eq. (6) for a laser wavelength of 1 μm at peak intensity of 1 x $10^{14}$ W/cm$^2$, initial plasma density of 1 x $10^{19}$ cm$^{-3}$ and electron temperature of 40 keV. The structures (c) honeycomb, (d) chains and (e) dimmer lattice were obtained using the (a) configuration and the (f) Kagome lattice by the (b) configuration.